# Peculiarities of electron transport and resistive switching in point contacts on TiSe$_2$, TiSeS and Cu$_x$TiSe$_2$

D. L. Bashlakov[1], O. E. Kvitnitskaya[1], S. Aswartham[2], Y. Shemerliuk[2], H. Berger[3], D. V. Efremov[2], B. Büchner[3,4], Yu. G. Naidyuk[1]

[1]B. Verkin Institute for Low Temperature Physics and Engineering, NAS of Ukraine, 61103 Kharkiv, Ukraine

[2]Leibniz Institute for Solid State and Materials Research, IFW Dresden, 01069 Dresden, Germany

[3]Ecole Polytechnique Federale de Lausanne, LPMC, CH-1015 Lausanne, Switzerland

[4] Institute of Solid State and Materials Physics and Würzburg-Dresden Cluster of Excellence ct.qmat, Technische Universität Dresden, 01062 Dresden, Germany

**Abstract**

TiSe$_2$ has received much attention among the transition metals chalcogenides because of its thrilling physical properties concerning atypical resistivity behavior, emerging of charge density wave (CDW) state, induced superconductivity etc. Here, we report discovery of new feature of TiSe$_2$, namely, observation of resistive switching in voltage biased point contacts (PCs) based on TiSe$_2$ and its derivatives doped by S and Cu (TiSeS, Cu$_x$TiSe$_2$). The switching is taking place between a low resistive mainly "metallic-type" state and a high resistive "semiconducting-type" state by applying bias voltage (usually <0.5V), while reverse switching takes place by applying voltage of opposite polarity (usually <0.5V). The difference in resistance between these two states can reach up to two orders of magnitude at the room temperature. The origin of the effect can be attributed to the variation of stoichiometry in PC core due to drift/displacement of Se/Ti vacancies under high electric field. Additionally, we demonstrated, that heating takes place in PC core, which can facilitate the electric field induced effect. At the same time, we did not found any evidence for CDW spectral features in our PC spectra for TiSe$_2$. The observed resistive switching allows to propose TiSe$_2$ and their derivatives as the promising materials, e.g., for non-volatile resistive random access memory (ReRAM) engineering.

**Intoduction**

Transition metal dichalcogenides (TMDs) are emerging materials, which, depending on their chemical compositions and crystal structure, can be in metallic, semimetallic, semiconducting, magnetic or even in superconducting state [Wilson]. Layered structure of TMDs with weak van der Waals bounding allows exfoliate these materials up to atomic thickness, which opens incredible potential for their applications in nanoelectronics, optoelectronics, spintronics, storage etc. [AKB]. Among TMDs, semimetallic $TiSe_2$ has received much attention because of its diverse and intriguing physical properties. First of all, it concerns anomaly in resistivity $\rho(T)$ around 160 K [Benda74], which evolved into a huge maximum for single crystals [Salvo76], rarely observed in metallic systems. Recently, Refs. [Huang17, Moya19] reported, that electronic transport properties of $TiSe_2$ are very sensitive to stoichiometry of crystals. For example, the maximum in resistivity is more pronounced for polycrystals with Se vacancies, while samples with improved stoichiometry shows a typical semiconducting behavior, with only an insignificant trace of $\rho(T)$ peak anomaly near 200 K [Huang17].

Another feature of $TiSe_2$ is second-order phase transition, which occurs near 200 K and results in doubling of the hexagonal *a* and *c* axes and superlattice formation [Salvo76, Salvo78]. Origin of this transition is still debated starting from early papers [Salvo76, Woo79] and discussions are focused on charge density wave (CDW) representation. At the same time, some recent paper [Watson19] explains "anomalous peak in the resistivity" due to "a crossover between a low-temperature regime with electron-like carriers only, to a regime around room temperature where thermally activated and highly mobile hole-like carriers dominate the conductivity". Spera et al. [Spera20], combining density functional theory and STM data analysis, concluded that CDW formation involves primarily electronic states away from the Fermi level deep inside the valence band. In this case, it is hardly to expect, that CDW transition will have impact on electronic transport properties of $TiSe_2$. Even this short overview demonstrates that $TiSe_2$ is a very interesting material, which can hide many more surprises. This was the reason to investigate $TiSe_2$ and related compounds by Yanson point-contact (PC) spectroscopy method [PCbook]. Recently, we applied this method to study other TMDs, such as $MoTe_2$, $WTe_2$ and $TaMeTe_4$ (*Me*= Ru, Rh, Ir), where electron-quasiparticle interaction [Naid19] and PC enhanced superconductivity were observed [Naid18] as well resistive switching in PCs with these TMDs was discovered [Naid21].

**Experimental details**

*Samples*. The single-crystals $TiSe_2$ and $1T-CuTiSe_2$ were grown by a conventional chemical vapor transport method [Levy83, Shemerliuk22]. More details of TiSeS single crystals growth and characterization are descried in Ref [Shemerliuk22].

*Point contact spectroscopy.* PCs were prepared by touching of a thin Ag or Au wire to a cleaved at room temperature flat surface of needle-like single-crystal flake or contacting its edge/side by this wire. Also, so-called "soft" PCs were made by dripping of a small drop of silver paint onto the cleaved sample surface/edge. The latter type of PCs demonstrates better stability versus temperature change. We create PCs one by one at low or room temperature inside the cryostat, and then sweep voltage back and forth with increasing of its amplitude until resistance switching was observed. Thus, we conducted resistive measurements on the hetero-contacts between a normal metal (Ag, Au or silver paint) and investigated samples. We measured current–voltage *I–V* characteristics of PCs and their first derivatives *dV/dI(V)*. The first derivative or differential resistance $dV/dI(V) \equiv R(V)$ was recorded by scanning the *dc* current *I* on which a small *ac* current *i* was superimposed using a standard lock-in technique. The measurements were performed in the temperature range from liquid helium up to the room temperature and at magnetic field up to 15T.

**Experimental results.**

Fig. 1(a) shows *dV/dI(V)* of PC on $TiSe_2$ with two symmetrically located maxima. Interesting that the *dV/dI(V)* bears a resemblance with resistivity *ρ(T)*, shown in the inset. Fig. 1(b) displays that the double maximum structure in *dV/dI(V)* narrows with increasing temperature and disappears above 170 K (see Fig. 1(b) inset), where maximum in *ρ(T)* is located. These observations testify in favor of the thermal regime in PC, when PC temperature increases with a bias voltage [Verkin79]. Fig. 2(a) reveals calculated asymmetric part of *dV/dI(V)*$^{as}$ for 3 different PCs. Here similar behavior is seen between *dV/dI(V)*$^{as}$ and Seebeck coefficient *S(T)*. This observation is also in support of the thermal regime in PC in the case of heterocontact created between two metals with a big difference in Seebeck coefficients [Naid85]. So, thermal transport prevails in PC, that is, temperature in PC ($T_{PC}$) rises with a bias voltage and in the case of fulfillment of Wiedeman-Franz law [Verkin79]:

$$T^2_{PC} = T^2_{bath} + V^2/4L, \qquad (1)$$

where $T_{bath}$ is the temperature of environment, $L = \pi^2 k_B^2/3e^2$ is the Lorenz number. At high voltage $eV \gg k_B T_{bath}$ or low temperature $T_{bath}$, Eq.(1) reduces to $eV = 2e\sqrt{L}T_{PC} = 2\pi/\sqrt{3}\ k_B T_{PC} = 3.63 k_B T_{PC}$. So, $T_{PC}$ rises linearly with a voltage with a rate 3.2 K/mV for standard Lorenz number $L_0 = 2.45 \cdot 10^{-8}$ $V^2/K^2$. In the case of a thermal regime *I-V* curve of PC can be calculated from the equation [Verkin79]:

$$I(V) = Vd \int_0^1 \frac{dx}{\rho(T_{PC}(1-x^2)^{1/2})}, \qquad (2)$$

where $d$ is the PC diameter. We applied Eq. (2) to calculate $dV/dI(V)$. We used resistivity $\rho(T)$ in Eq. (2) from the inset in Fig.1(a). To fit position of the maximum in $dV/dI(V)$, we increased Lorenz number, to fit amplitude of the maximum, we added residual resistivity $\rho_0$ to $\rho(T)$ and to receive correct zero-bias resistance of PC, we variate diameter. As a result, theoretical curve in Fig. 2(b) fits perfect the experimental $dV/dI(V)$. Thus, we confirmed thermal regime even quantitatively.

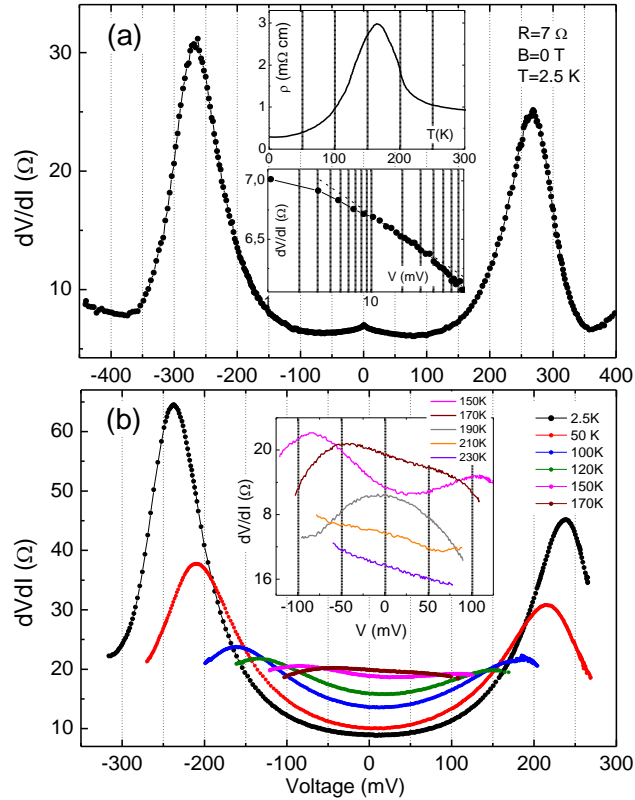

Fig.1. (a) *dV/dI(V)* of PC on $TiSe_2$ – Ag. Insets show typical behavior of resistivity of $TiSe_2$ [Salvo78] with pronounced maximum and *dV/dI(V)* around zero-bias in log-scale, respectively. (b) Behavior of *dV/dI(V)* of another PC on $TiSe_2$ at different temperatures.

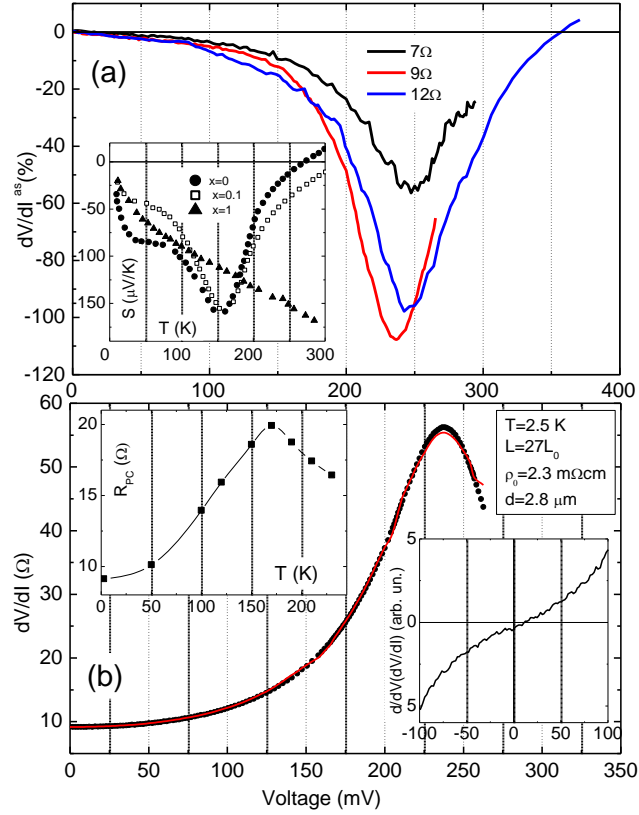

Fig.2. (a) Calculated asymmetric part $dV/dI^{as}=[dV/dI(V)- dV/dI(-V)]/2dV/dI(V=0)$ for 3 PCs with resistance 7, 9 and 12 Ω. Left inset shows Zeebeck coefficient for $TiSe_{2-x}S_x$ for x=0, 0.1 and 1 [Lopez87]. (b) Calculation of $dV/dI(V)$ (solid red line) in the thermal regime according to Eq. (2) in comparison with symmetrized experimental $dV/dI(V)$ at 2.5K from Fig. 1(b) (black symbols). Fit parameters $\rho_0$, $d$ and $L$ are shown in the corner. Left inset shows dependence of PC resistance at zero bias for the PC from Fig. 1(b). Right inset shows digital derivative of $dV/dI$ at 2.5 K from Fig. 1(b).

Important observation from temperature dependence of $dV/dI(V)$ demonstrates inset in Fig. 2(b), where zero-bias PC resistance $R_{PC}(T)$ versus temperature is shown. We see that $R_{PC}(T)$ resembles $\rho(T)$ behavior. This is in line with the Wexler formula [Wexler66, PCSbook] for $R_{PC}(T)$, which contains the sum of ballistic Sharvin (temperature independent) and diffusive Maxwell resistance (temperature dependent):

$$R_{PC} \approx 16\rho l/3\pi d^2 + \rho(T)/d, \qquad (3)$$

where $l$ is the electron mean free path, $\rho l = p_F/ne^2 \approx 1.3 \times 10^4 n^{-2/3} \approx 5 \times 10^{-10}$ Ω cm$^2$, if we use carrier density, e.g., $n \approx 5 \times 10^{20}$ cm$^{-3}$ from [Salvo78]. Such similarity between $R_{PC}(T)$ and $\rho(T)$ behavior testify, that material in PC core corresponds to $TiSe_2$, not to some disturbed substance, oxide, contaminated surface etc. Here, it is necessary to add, that using Eq. (3) for determination of PC diameter $d$ must be taken with a care because residual resistivity $\rho_0$ in PC core is unknown. Therefore to fit $dV/dI(V)$ in Fig. 2(b), we used $\rho_0$ as a fit parameter.

Fig. 3(a) displays observation of resistive switching in PC. When voltage reaches about -300 mV, *dV/dI(V)* jumps above -600 mV and then by sweeping to positive polarity *dV/dI(V)* reveals sharp zero-bias peak instead of broad minimum at initial state. Then at positive voltage around 300 mV, *dV/dI(V)* comes back to the low resistive state (LRS). Three such cycles with switching between LRS and high resistance state (HRS) is shown in Fig. 3(a). Fig. 3(b) displays resistive switching for another PC at different temperatures. It can be seen that the switching is preserved up to the room temperature, although PC is less stable at high temperatures. Note, that the pronounced sharp peak for HRS has distinctly log-behavior.

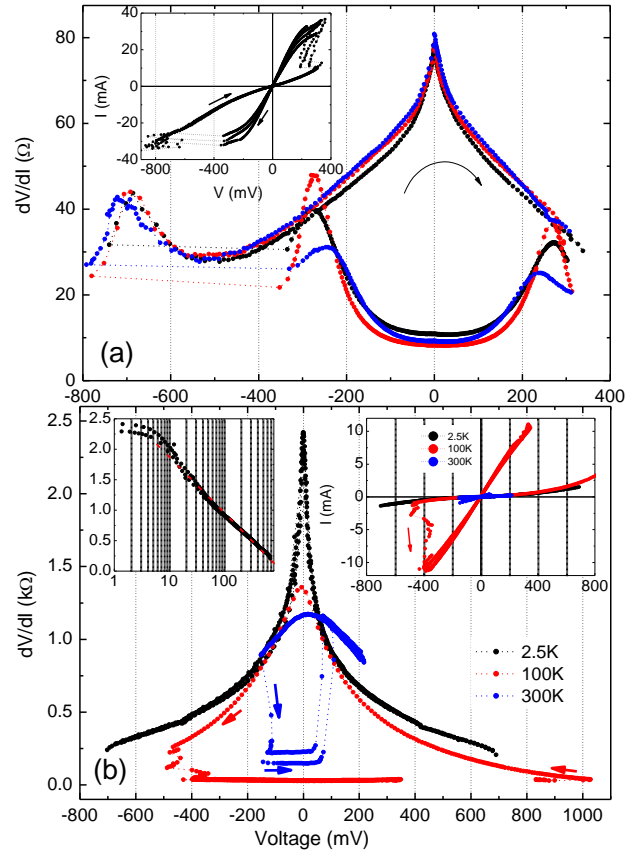

Fig. 3. (a) Resistive switching in *dV/dI(V)* for $TiSe_2$ – Ag PC by „clock wise" sweeping of bias voltage for 3 cycles. Inset shows *I-V* curves for the same PC. (b) Resistive switching in *dV/dI(V)* for $TiSe_2$ PC at 3 different temperatures. Left insert shows *dV/dI(V)* at 2.5K in logarithmic scale. Right inset shows *I-V* curves for PC from the main panel.

Fig. 4 demonstrates observation of similar switching effect for TiSeS compound, where half of Se is substituted by S. As shown in [Lopez87], prounounced maximum in *ρ(T)* reduces by substitution Se by S and is absent for TiSeS. Nevertheless, the switching in TiSeS PCs is also observed at critical voltage close to 300 mV and *dV/dI(V)* in HRS has also log-behavior. LRS exibits broad minimum, although in some cases small zero-bias maximum is present, e.g., as in Fig. 4(b) and Fig.1(a). As can be seen from the inset in Fig. 4(b), a magnetic field up to 12 T has no visible effect on this peak. Also switching was observed in magnetic field up to 15 T.

Finally, we studied $TiSe_2$ intercalated by Cu, where superconducting state evolves below 3.5K. For this compound, we again observed switching in *I-V* curves and their derivative (see Fig. 5). In this case, we measured switching effect in more details, that is for two main crystallografic directions along and perpendicular to the *c*-axis taking into account large anisotropy with up to two order of magnitude larger resistivity along the *c*-axis as compared to the basal plane [Wu07].

However, the switching was observed for both directions. The only difference is mentioned that for the *c*-axis or for out of plain direction, the zero-bias peak was presented regularly for LRS, while for in-plain direction there was little or no zero-bias peak and *dV/dI* shows smoth minimum similar to that shown in Fig. 3(a). Fig.5 demonstrates also interesting temperature behavior of *dV/dI*. Namely, direction of switching is reversed above 200K for PC from panel (a). Another observation is counterintuitive increase of switching amplitude for PC from panel (b) above 200K, where we have observed record high change of the resistance up to two order of magnitude at the room temperature. We must also underline, that reproducibility of the switching loops is lower by increasing of temperature, especially at the room temperature.

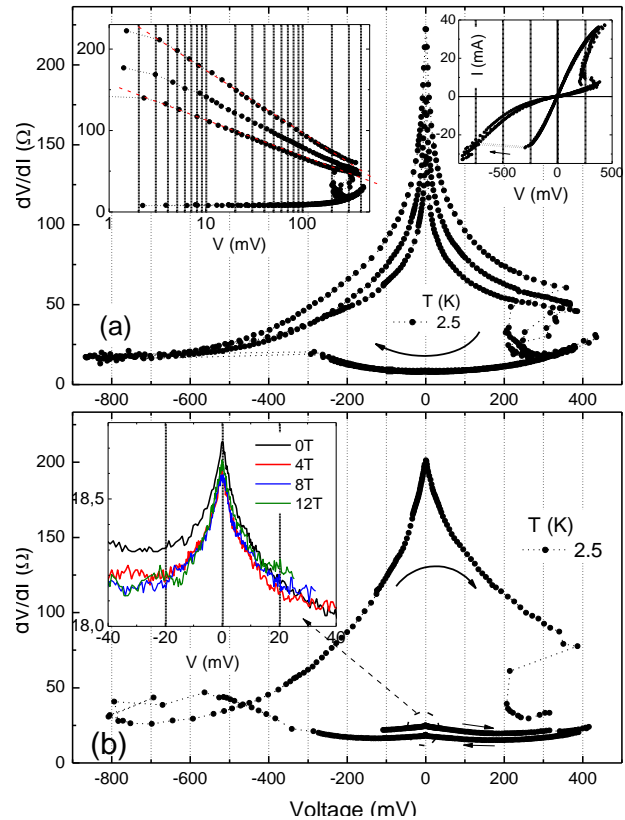

Fig.4. (a) Resistive switching in *dV/dI(V)* for TiSeS – Ag PC by „clock wise" sweeping of bias voltage for 3 cycles. Right inset shows *I-V* curves for the same PC. Left insert shows *dV/dI(V)* at 2.5K in logarithmic scale. (b) Resistive switching in *dV/dI(V)* for another TiSeS PC at 2.5K. Left inset shows *dV/dI(V)* for the same PC at low resistance state for different magnetic fields.

Regarding statistics. We measured 12 switching loops for $TiSe_2$ PCs with resistance in LRS between 5 and 33 Ω and 24 loops for TiSeS PCs with resistance between 6 and 50 Ω, all at helium temperature. Switching from LRS to HRS for vast majority of PCs (an exception shown in Fig. 3(b)) was at negative polarity (positive polarity on samples), while backward HRS to LRS switching was observed at positive polarity. LRS to HRS switching occurs in the range between -200 and -400 mV with average value around -280 mV for both compounds, while HRS to LRS switching is taking place also within +200 and +400 mV with average value around +280 mV for $TiSe_2$ and around +330 mV for TiSeS.

We measured the swithching voltages for 38 $Cu_xTiSe_2$ PCs. Statistical data are presented in Fig.6 as a histogram for in plane and out of plane direction. Here, the switching also occurs preferably in the range between 200 and 400 mV for both polarities. The most visible effect is that the polarity of the switching, e.g. for LRS to HRS, is equally distributed between positive and negative voltages for out of plane (c-axis) direction. But, for in plane (PC to edge/side) direction only two PCs show LRS to HRS transition at positive voltage, while other 16 PCs at negative polarity.

**Discussion.**

Characteristic maximum in *dV/dI(V)* similar to that in *ρ(T)* along with the calculation of *dV/dI(V)* in the thermal regime provides direct evidence of PC heating. Thermal regime is opposite to the ballistic or diffusive regime, when spectroscopic information can be obtained from the second derivative of *I-V* curve by Yanson PC spectroscopy [PCSbook]*,* e.g., as to electron-phonon interaction. As we see from derivative of *dV/dI* curve in Fig. 2(b) right inset (it corresponds to the second derivative of *I-V* curve), no spectral features are visible in the region up to +/- 100 mV, where typical excitations like phonons and/or may be CDW gap are expected to be seen. Note, that the maximal phonon energy in $TiSe_2$ is 40 meV [Calandra11], CDW transition temperature is about 200 K (or about 20 meV) [Salvo76] and estimated CDW gap width is 70 meV [Spera20]. The reason for the absence of the spectral features can be small inelastic mean free path of electrons in comparison with the PC size. E.g., PC size for spectral regime must be not larger than about 100 nm for simple metals [PCSbook], while, as follows from the calculation in Fig. 2(b), PC diameter is more than an order of magnitude larger. Thus, so far, spectral information as to phonons or any evidence of CDW state is not detected in our PC measurements.

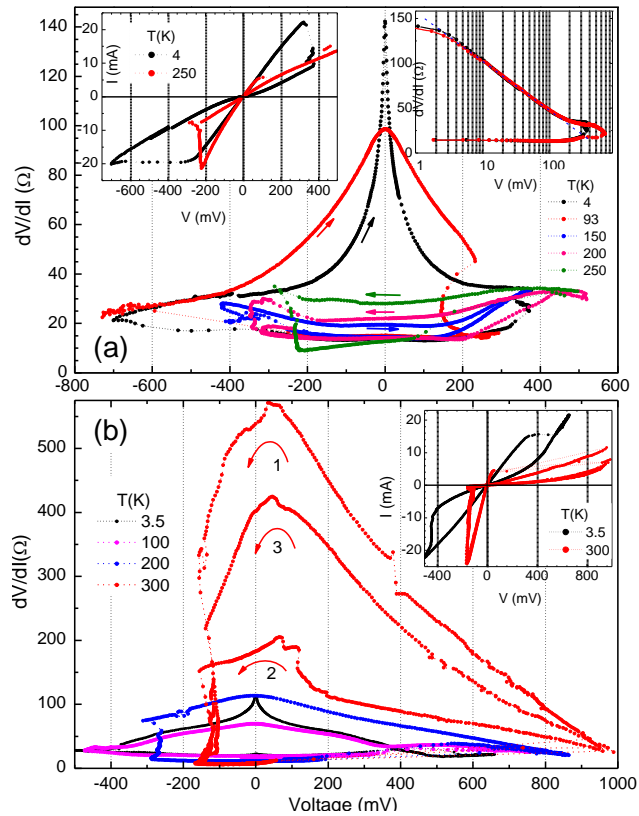

Fig.5. (a) Resistive switching in *dV/dI(V)* for $Cu_xTiSe_2$ – Ag PC at different temperatures. Right insert shows *dV/dI(V)* at 4K for both polarities in log-scale. Left inset shows *I-V* curves for PC from the main panel at helium and room temperatures. (b). Resistive switching in *dV/dI(V)* for another $Cu_xTiSe$ PC by „anticlockwise" sweeping of the bias voltage. Three switching cycles are shown at room temperature. Right inset shows *I-V* curves for the same PC at helium and room temperatures.

Draws attention to itself, the large Lorenz number, which is used to fit the position of the maximum in the calculated *dV/dI* with the experimental curve. To clarify this point, we calculated Lorenz number as $L=k\rho/T \approx 3.6 \cdot 10^{-6}$ $V^2/K^2$, using resistivity data $\rho \approx 0.3$ mΩcm and thermal conductivity $k \approx 0.3$ W/cm K at 25 K from Ref. [Ayache81]. As a result, the estimated Lorenz number is two order of magnitude larger than the standard Lorenz number $L_0=2.45\cdot 10^{-8}$ V$^2$/K$^2$. This agrees with our data as to high Lorenz number and it means that the non electron contribution prevails in the thermal conductivity of TiSe$_2$.

Some PCs, e. g., as shown in Fig. 1(a) and Fig. 4(b) inset, displays shallow zero-bias maximum in LRS with close to log-behavior. Similar log-increase of resistivity upon cooling was observed for the polycrystalline TiSe$_2$ samples [Moya19], what was connected to weak-localization effects due to low dimensionality. However in our case, magnetic field has no influence on this zero-bias maximum up to 12 T, as it is shown, e.g., for TiSeS PC in Fig. 4(b), inset. Other processes, which can be responsible for zero-bias maximum in PC spectra, can be Kondo effect or two-level systems [PCSbook], characteristic for a disordered state. Among these two effects, only the latter is robust as to a magnetic field. Of course, some disorder of the lattice cannot be ruled out in a PC.

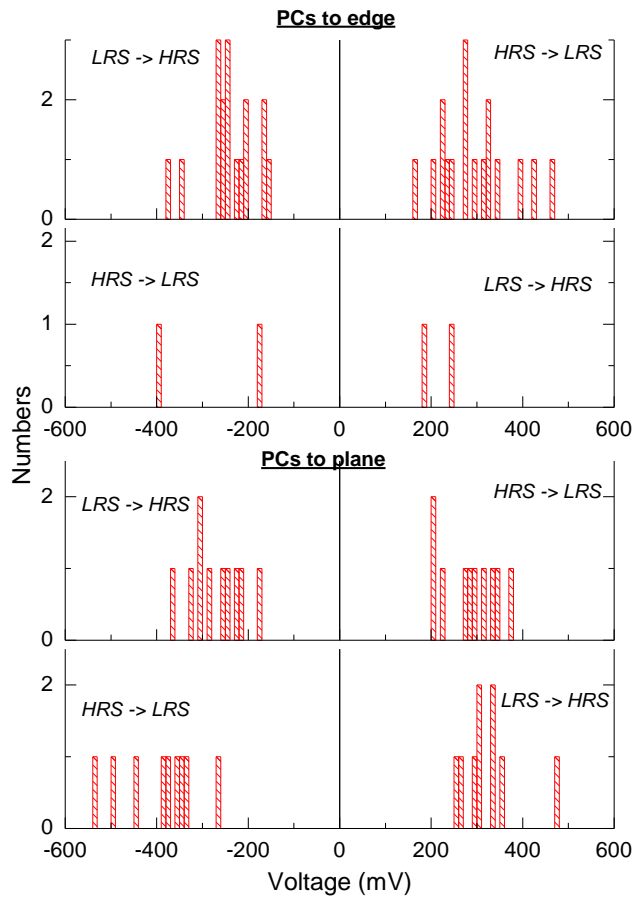

Fig.5. Histogramm of switching voltages in *dV/dI(V)* for 38 PCs Cu$_x$TiSe$_2$ PC at helium temperatures prepared for in plane and out of plane (edge/side) geometry.

Let's move on to the discovered switching effect. At first, we observed this effect in TiSe$_2$, but then also for TiSeS and Cu$_x$TiSe$_2$, where CDW state is absent. That is, the CDW state, if any, does not affect this phenomenon in any way. Second, while LRS has mainly metallic-like behavior, HRS displays nonmetallic semiconducting-like behavior. Futhermore , *dV/dI* in HRS has log-dependence as show insets in Fig. 3(b), Fig. 4(a) and Fig. 5(a). Interesting, that similar was observed in the transition metals ditellurides [Naid21]. It looks like a small zero-bias maximum for LRS, as display Figs. 1(a) and 4(b), evolves to the huge one in HRS. We should to note, that as follows from Ref. [Huang17], Se vacancy defects have exceptional influence on TiSe$_{2-\delta}$ properties. According to [Huang17]: "the most pronounced *ρ(T)* CDW anomaly is confirmed to occur when *δ* is near ~0.12, instead of being stoichiometric", while samples with *δ* ~ 0.08 demonstrate monotonous *semiconducting-like* behavior with missing maximum in *ρ(T)*. Thus, the origin of the switching effect may be due to a change of stoichiometry in PC core, that is because of formation of domain in PC with different structure, due to drift/displacement of Se vacancies under a high electric field. According to [Huang17], activation energy of vacancies generation is about 100 meV, that is lower than the switching voltage, what is in support of the movement of vacancies. In this case, the heating of PC core due to the thermal regime, which we mentioned at the start of the discussion, can facilitate the switching effect. We can estimate the temperature of PC at which the switching takes place. As it is seen from Fig. 3(a), the switching occurs just after the maximum in *dV/dI(V)*, that is the temperature is not a much higher than the temperature of the maximum in *ρ(T)*, that is about 200K. This avoids formation of some conductive filament due to, e.g., melting of the contacted materials in PC core.

As shows histograms in Fig.5, switching, e.g. from LRS to HRS, can be observed for both polarities of voltage. Then we must assume that there are vacancies with opposite charge to Se vacancies. In this case only Ti vacancies can play such a role.

It would be also interesting to figure out contribution of domain wall in the observed switching effect. Especially, what is the nature of observed logarithmic behavior in *dV/dI(V)* at HRS.

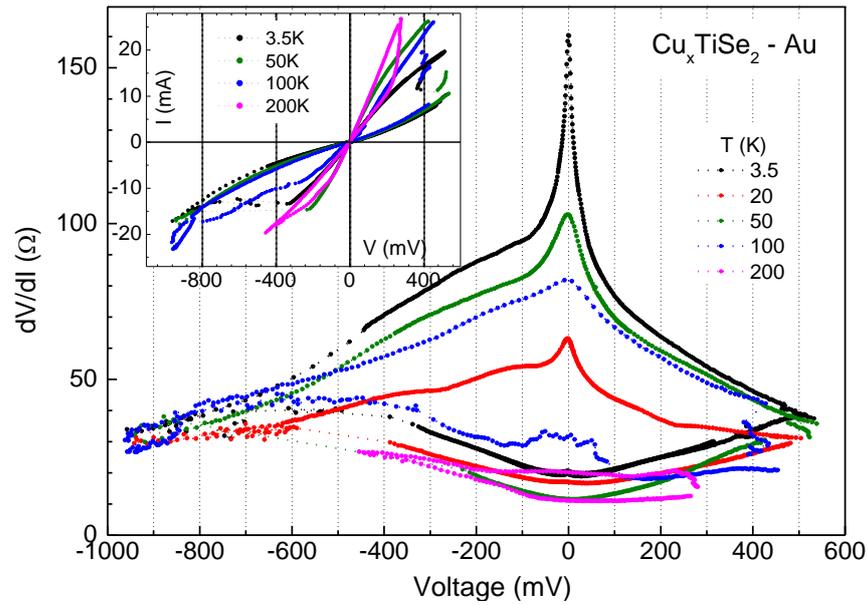

Fig.6. (a) Resistive switching in *dV/dI(V)* for $Cu_xTiSe_2$ – Au PC at different temperatures. Inset shows *I-V* curves for PC from the main panel.

It should be noted that electrochemically active Ag is widely used as an electrode material for the construction of a cell with resistive switching. In this case, the formation of conductive Ag dendrites can occur due to an electrochemical process [Waser09]. To test this possibility, we additionally used an Au counter electrode, which is electrochemically inert. Nevertheless, we also observed a similar switching effect as in the case of Ag (see Fig.6.), what rejects the explanation of the observed switching due to the growth of Ag filaments.

In conclusion, we did not found any clear evidence for CDW spectral features presence in our $TiSe_2$ PC spectra, instead, we demonstrate that the thermal regime is implemented in PCs, when *dV/dI(V)* behavior is determined by the resistivity *ρ(T)*. We observed resistive switching in $TiSe_2$, as well as in TiSeS and $Cu_xTiSe_2$ PCs, between metallic-like LRS and semiconducting-like HRS with changing resistance up to two orders of magnitude at the room temperature. Therefore, discovering of the resistive switching add TiTe$_2$, TiTeS and Cu$_x$TiSe$_2$ to the list of compounds which can be promising, e.g., to non-volatile ReRAM engineering actual, e.g., for neuromorphic computing architectures and Internet of Things [Sang20]. On the other hand, this paper demonstrates that PCs, which form the basis of Yanson PC spectroscopy [PCbook], can be successfully used as a simple technique to discover new functionalities of already well known and emergent materials.

**Acknowledgement**

We are grateful to S. Gaβ and A. Wolter-Giraud for the technical assistance, S. Borisenko for helpful conversations. We would like to acknowledge funding by Alexander von Humboldt and Volkswagen Foundation. DB, YuN and OK are also grateful for support by the National Academy of Sciences of Ukraine under project Φ19-5 and thankful to the IFW Dresden for hospitality. SA appreciating support from Deutsche Forschungsgemeinschaft through Grant No: AS 523/4-1.